\documentclass[aps,pre,floatfix,superscriptaddress]{revtex4-1}
 \pdfoutput=1
 \usepackage[dvipsnames,rgb,dvips]{xcolor}
 \usepackage{graphicx}
 \usepackage{amsmath,amssymb}
 \usepackage{bm}
 \usepackage[T1]{fontenc}
 
 \newcommand{\dd}{{\rm d}}
 \newcommand{\tr}{^{\sf T}}
 \newcommand{\ve}[1]{\bm{#1}}
 \newcommand{\ma}[1]{\mathbf{#1}}

\begin{document}

\title{The emergence of the rescue effect from the interaction of local stochastic dynamics within a metapopulation}
\author{Anders Eriksson}
\email{aje44@cam.ac.uk}
\affiliation{Evolutionary Ecology Group, Department of Zoology, University of Cambridge, Cambridge CB2 3EJ, United Kingdom}
\affiliation{Integrative Systems Biology Laboratory, King Abdullah University of Science and Technology (KAUST), Thuwal 23955-6900, Kingdom of Saudi Arabia}
\author{Federico El\'ias-Wolff}
\affiliation{Department of Physics, University of Gothenburg, SE{-}41296 Gothenburg, Sweden}
\author{Bernhard Mehlig}
\email{bernhard.mehlig@physics.gu.se}
\affiliation{Department of Physics, University of Gothenburg, SE{-}41296 Gothenburg, Sweden}
\affiliation{Centre for Evolutionary Marine Biology, University of Gothenburg, Gothenburg, Sweden}
\author{Andrea Manica}
\email{am315@cam.ac.uk}
\affiliation{Evolutionary Ecology Group, Department of Zoology, University of Cambridge, Cambridge CB2 3EJ, United Kingdom}

\begin{abstract}
Immigration can rescue local populations from extinction, helping to stabilise a metapopulation. Local population dynamics is important for determining the strength of this rescue effect, but the mechanistic link between local demographic parameters and the rescue effect at the metapopulation level has received very little attention by modellers. We develop an analytical framework that allows us to describe the emergence of the rescue effect from interacting local stochastic dynamics. We show this framework to be applicable to a wide range of spatial scales, providing a powerful and convenient alternative to individual-based models for making predictions concerning the fate of metapopulations. We show that the rescue effect plays an important role in minimising the increase in local extinction probability associated with high demographic stochasticity, but its role is more limited in the case of high local environmental stochasticity of recruitment or survival. While most models postulate the rescue effect, our framework provides an explicit mechanistic link between local dynamics and the emergence of the rescue effect, and more generally the stability of the whole metapopulation.
\end{abstract}

\maketitle

\section*{Introduction}
\label{sec:intro}
The increasing rate of habitat destruction and fragmentation poses a major threat to the persistence of many species \citep{pavlacky2012,dornier2012}. Connectivity among populations has been long recognised to play a key role in preventing extinction \citep{hanski1999,macarthur1963}: immigration can lead to the recolonisation of previously extinct patches, thus promoting the long-term persistence of the network of populations, i.e. the metapopulation \citep{hanski1999}. Also in the absence of population turnover, immigration can reduce the probability of local extinction by boosting numbers of inhabitants under adverse conditions, thus buffering local populations against demographic and environmental stochasticity. This reduction in local extinction probability was termed the ``rescue effect'' by \citet{brown1977}, and ecologists have accumulated a wealth of empirical evidence of its potential importance \citep[e.g.][]{pavlacky2012,dornier2012,lawson2012,hanski1998}.

While it is intuitively obvious that the rescue effect can play a pivotal role in determining the persistence of local populations as well as of the entire metapopulation, quantifying the link between local dynamics and the strength of the rescue effect is a challenging endeavour. The classical Levins' metapopulation model \citep{levins1969} considers patches as either occupied or empty, thus ignoring local population dynamics within patches. In an attempt to integrate the rescue effect in Levins' model, several authors have assumed that the extinction probability of a given patch is a function of the number of occupied patches \citep{hanski1993,gotelli1993}. However, the shape of this function cannot be derived from first principles, and has to be fitted to the data for each species of interest. Thus, the lessons learnt from this approach are specific to the dataset in question.

In order to investigate the factors that affect the strength of the rescue effect it is necessary to explicitly model the local stochastic population dynamics. Agent-based models can be used for this purpose \citep[e.g.][]{keeling2002}, but it is often difficult to understand the mechanisms that determine the global metapopulation dynamics, and to quantify the influence of the different factors characterising the local population demography. 

An analytical treatment of the metapopulation is much more desirable and informative, but it is not trivial. \citet{lande1998} were the first to develop such an analytical model based on a diffusion approximation for the local population dynamics. Their approach characterises the metapopulation dynamics as a sequence of isolated, local extinction and recolonisation events, and relies on the assumption of very low migration rates. This assumption is appropriate when a large fraction of suitable patches is not occupied near demographic equilibrium. This scenario is at odds with the observation that the majority of suitable patches tend to remain occupied in many metapopulations \citep[e.g.][]{andrewartha1954,kitching1971,kaitala1987,harrison1995,solbreck1991}. \citet{keeling2000} took a different approach, relying on a moment-closure technique, to investigate the persistence of stochastic metapopulations. This approach relies on knowing the distribution of population sizes, a problematic restriction if we are interested in investigating the dynamics of a metapopulation close to extinction, for which the shape of the distribution is likely to differ from the corresponding shape near demographic equilibrium.

In this paper, we develop an analytical framework, based on the identification of the slow and fast modes underlying the metapopulation dynamics. Our approach allows us to investigate the link between local stochastic dynamics and the strength of the rescue effect. We compare our results to spatially explicit simulations to explore the spatial scales at which our analytical results are relevant. While for this paper we focus on a model that describes the local dynamics in terms of a Ricker map \citep{ricker1954}, the approach can be applied to other models for population dynamics, formulated in terms of discrete non-overlapping generations or in terms of continuously occurring birth and death processes.

\section*{Metapopulation Model}
\label{sec:metapopulation}

Our model considers $N$ patches connected by global dispersal, and we specifically focus on metapopulations consisting of a large number of patches.  Each patch can be inhabited by a population that evolves in discrete, non-overlapping generations. Each generation has two phases, the first corresponding to local recruitment and the second to dispersal to other populations. For biological realism, we use the model by \citet{melbourne2008} to represent local recruitment and density-dependent survival. This model is a stochastic version of the classical deterministic Ricker model \citet{ricker1954} that incorporates the effect of both demographic and environmental stochasticity within populations. 
In our model each individual has a Poisson-distributed number of progeny, with mean given by the recruitment rate $R_i$ for individual $i$. The individual recruitment rate $R_i$ is itself random (independent between generations), giving rise to demographic and environmental stochasticity. The environmental stochasticity determines a mean (environmental) recruitment rate, $R_{\rm E}$, for the local population ($R_{\rm E}$ is gamma distributed with mean $R$ and shape parameter $k_{\rm E}$). Given the environmental state of the local population, individual recruitment rates $R_i$ are independent gamma distributed stochastic variables with mean $R_{\rm E}$ and shape parameter $k_{\rm D}$ (giving rise to demographic recruitment stochasticity). Finally, each progeny survives (independently) to adulthood with density-dependent probability $\exp(-\alpha n)$, where $n$ is the number of adults in the population and $\alpha$ regulates the strength of the density dependence. As shown by \citet{melbourne2008}, the expected number of progeny surviving to adulthood is the same as in the classical Ricker dynamics (with recruitment $R$ and density dependence $\alpha$), but the variance of the distribution is strongly affected by the amount of demographic and environmental recruitment stochasticity (defined by the shape parameters $k_{\rm E}$ and $k_{\rm D}$). Putting it all together (following \citep{melbourne2008}) the probability of having, after the first phase, $j$ individuals in a population with $i$ adults ($P_{ij}$) can be written as
\begin{subequations}\label{eq:1}
\begin{equation}\label{eq:1a}
P_{ij} = \int_0^{\infty}\!\!\! \frac{k_{\rm E}}{R \Gamma(k_{\rm E})} 
\bigg(\frac{k_{\rm E} R_{\rm E}}{R} \bigg)^{k_{\rm E}-1} 
e^{\frac{k_{\rm E} R_{\rm E}}{R}} \binom{i+k_{\rm D}j-1}{k_{\rm D}j-1}
\bigg(\frac{R_{\rm E} e^{-\alpha j}}{k_{\rm D}+R_{\rm E} e^{-\alpha j}} \bigg)^i 
\bigg(\frac{k_{\rm D}}{k_{\rm D}+R_{\rm E} e^{-\alpha j}} \bigg)^{k_{\rm D}j} \dd R_{\rm E} \, .
\end{equation}
The original formulation by \citet{melbourne2008} detailed above describes a system where environmental stochasticity acts on recruitment. We also consider an alternative model in which demographic stochasticity of recruitment is the same as in Eq.~\eqref{eq:1a}, but where environmental stochasticity affects survival. Survival stochasticity is modelled by dividing  $\alpha$ by a gamma-distributed variable with unit mean and shape parameter $k_{\rm A}$. Thus, for this model we have
\begin{equation}\label{eq:1b}
P_{ij} = \int_0^{\infty}\!\!\! \frac{k_{\rm A}}{\Gamma(k_{\rm A})} 
\Big( k_{\rm A} z \Big)^{k_{\rm A}-1} 
e^{-k_{\rm A} z} \binom{i+k_{\rm D}j-1}{k_{\rm D}j-1}
\bigg(\frac{R e^{-\alpha j/z}}{k_{\rm D}+R e^{-\alpha j/z}} \bigg)^i 
\bigg(\frac{k_{\rm D}}{k_{\rm D}+R e^{-\alpha j /z}} \bigg)^{k_{\rm D}j} \dd z \, .
\end{equation}
\end{subequations}
In the following, the derivations do not rely on the specific form of $P_{ij}$ in Eqs.~\eqref{eq:1a} and \eqref{eq:1b}, and we will therefore use ``Eq.~\eqref{eq:1}'' when either Eq.~\eqref{eq:1a} or Eq.~\eqref{eq:1b} can be used interchangeably.

In the second phase, the dispersal phase, each surviving progeny disperses with probability $m$, and otherwise stays in the local population. For simplicity we assume that a dispersing individual is equally likely to move to any other patch; if the destination patch is currently unoccupied, the individual colonises it and starts a new local population. We later investigate the effect of the range of dispersal using spatially explicit simulations of the dynamics.

As a result of demographic and environmental stochasticity, individual populations undergo substantial fluctuations in size (Fig.~\ref{fig:1}). When the number of populations is large, by contrast, we expect that fluctuations in local populations have little effect upon the global distribution of population sizes in the metapopulation. In this limit of many patches, we formulate the metapopulation dynamics in terms of the fraction $f_i$ of populations with $i$ individuals. The fraction $f_i$ after local recruitment and density dependence (denoted as $f_i^{\rm R}$) can be written as
\begin{equation}\label{eq:2}
f_i^{\rm R}(t) = \sum_{i=0}^{\infty} P_{ij}f_j(t) \, ,
\end{equation}
where $P_{ij}$ is given by Eq.~\eqref{eq:1}. After dispersing individuals have left their natal populations, but before they have arrived to their new populations, the fraction $f_i^{\rm D}$ is given by
\begin{equation}\label{eq:3}
f_i^{\rm D}(t) = \sum_{j=i}^{\infty} \binom{j}{j-i} (1-m)^i m^{j-i} f^{\rm R}_j(t) \, .
\end{equation}
Finally, as there is no mortality during dispersal (mortality only occurs within the natal patch in our model), the mean number of individuals arriving to each patch is equal to the mean number of individuals dispersing out of each patch. Because of the large number of patches, the distribution of individuals arriving to each patch is approximately Poisson distributed, allowing us to write the dispersal rate (i.e. the expected number of individuals arriving to a patch), $I$, as:%
\begin{equation}\label{eq:4}
I(t) = \sum_{i=0}^{\infty}m i f_t^{\rm R}(t) \, .
\end{equation}
We obtain the population-size distribution for the next generation as:
\begin{equation}\label{eq:5}
f_i(t+1) = \sum_{j=0}^i \frac{I(t)^j}{j!} e^{-I(t) j} f_{j-i}^{\rm D}(t) \, .
\end{equation}
Eqs. (\ref{eq:1}-\ref{eq:5}) can be written as a single equation for the dynamics in matrix form
\begin{equation}\label{eq:6}
\Delta \ve{f}(t) = \ve{f}(t+1)-\ve{f}(t) = \ma{A}(I(t))\ve{f}(t)\, ,
\end{equation}
where $\ve f$ is the vector of population size frequencies ($f_i$) and $\ma{A}$ is a matrix that depends on the dispersal rate $I(t)$.

Note that Eqs.~(\ref{eq:1}-\ref{eq:4}) describe processes that occur independently in each patch; only in the dispersal stage [Eq.~\eqref{eq:5}] have we taken advantage of the assumption that the number of patches is large and treated $f_i(t)$ as deterministic rather than stochastic. As we will show later using simulations, this approximation is well justified. We also point out that, because $I(t)$ depends upon $f_i(t)$ in Eq.~\eqref{eq:4}, the left-hand side of Eq.~\eqref{eq:6}, $\Delta\ve{f}(t)$, depends nonlinearly on the distribution of population sizes in generation $t$. This non-linearity is important in determining how we can analyse the behaviour of the system in the following section. 

\subsection*{Spatial scale relevant for our spatially implicit model}
\label{sec:Spatial}

To test the importance of our assumption of a common dispersal pool, we compared the predictions of our spatially implicit model to spatially explicit simulations of the stochastic model on a two-dimensional grid of local populations connected by dispersal, varying the range of dispersal to explore the spatial scale at which the analytical framework is relevant (see \ref{app:A} in the electronic supplementary material (ESM) for details). We quantified the behaviour of the metapopulation by estimating steady-state mean population sizes for a range of dispersal probabilities. The results of our analytical framework were in excellent agreement with the simulations (Fig.~\ref{fig:2}), except for dispersal restricted to the nearest neighbour population in the lattice (yellow triangles in Fig.~\ref{fig:2}). Even at this most local form of dispersal, the analytical framework was able to capture qualitatively the relationship between our observable (mean population size) and the probability of dispersal $m$.

\section*{Dynamical analysis}
\subsection*{Identifying the slow mode underlying the metapopulation dynamics}

\begin{figure}[tp]
   \includegraphics[width=0.9\textwidth]{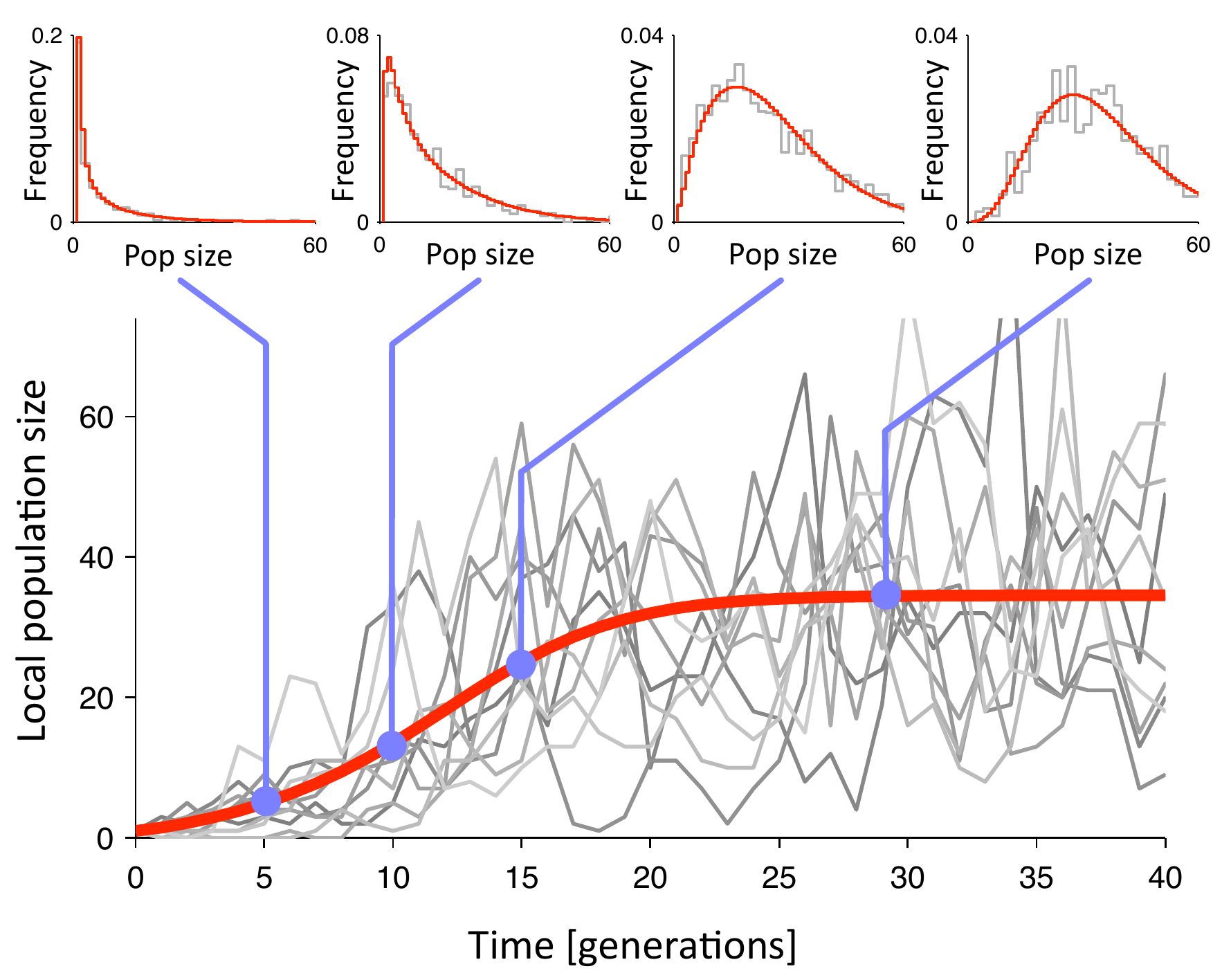}
   \caption{\label{fig:1}  
Trajectories of individual local population sizes through time are highly variable (grey lines, only a few shown out of 400 patches). However, the distribution of population sizes evolves smoothly and predictably (insets, grey lines are simulation and the red lines is the prediction from the deterministic approximation), and mean population size progressively increases and eventually converges to a stable level (red line, deterministic approximation). Parameters are: $R=1.5$, $\alpha=0.01$, $m=0.1$, $k_{\rm D}=1$, $k_{\rm E}=10$ (environmental recruitment stochasticity).}
\end{figure}

As mentioned above, individual populations exhibit substantial population-size fluctuations (Fig.~\ref{fig:1}). But the distribution of population sizes across the whole metapopulation (insets in Fig.~\ref{fig:1}) changes relatively slowly through time since demographic events in individual populations have very little effect as long as the number of populations is large, indicating a separation of time scale between local and global dynamics and the existence of a slow mode. We also find that, when the metapopulation recovers from different population sizes, it retraces the same trajectory to the global stationary state (except for brief transients, lasting a few generations after the perturbation, see Fig.~\ref{fig:S1} in the ESM). This separation of time scales is a common property of large interconnected non-linear systems, and there are well-established methods for analysing the fast and slow components of such systems, see for example Ref. \citep{gardiner2004}.

In general, fast and slow components of dynamical systems are found by analysing the eigenvectors and eigenvalues of the linearised dynamics. Eigenvalues of small magnitude correspond to slow components. In our case the structure of the dynamics is determined by the dispersal rate $I$ [Eq.~\eqref{eq:6}]. Therefore this quantity is a natural observable for parameterising the slow mode. Dispersal between patches renders the metapopulation dynamics nonlinear (the $I$-dependence of $\ma{A}$ makes Eq.~\eqref{eq:6} nonlinear). Therefore the eigenvectors and eigenvalues of the linearised dynamics (determined by the structure of the Jacobian) depend on the state of the metapopulation (e.g. the degree of occupancy, its proximity to extinction, etc.). To accommodate this important fact, we have to solve locally with respect to $I$ for the fast and slow components, and determine how they interact (see \ref{app:B} in the ESM).

Numerically, the slow mode corresponding to a given rate of dispersal $I$ (i.e. the local solution) can be found by spectral decomposition of the linearised dynamics (see \ref{app:B} in the ESM for details). The distribution of population sizes must (a) be normalised such that it sums to unity, and (b) have a rate of dispersal that matches the given value of $I$ (two linear constraints). To uniquely determine the slow mode, we use the adiabatic approximation that the population sizes distribution is unchanged along the directions of the fast degrees of freedom (as determined from the spectral decomposition). 

\begin{figure}[tp]
   \includegraphics[width=0.6\textwidth]{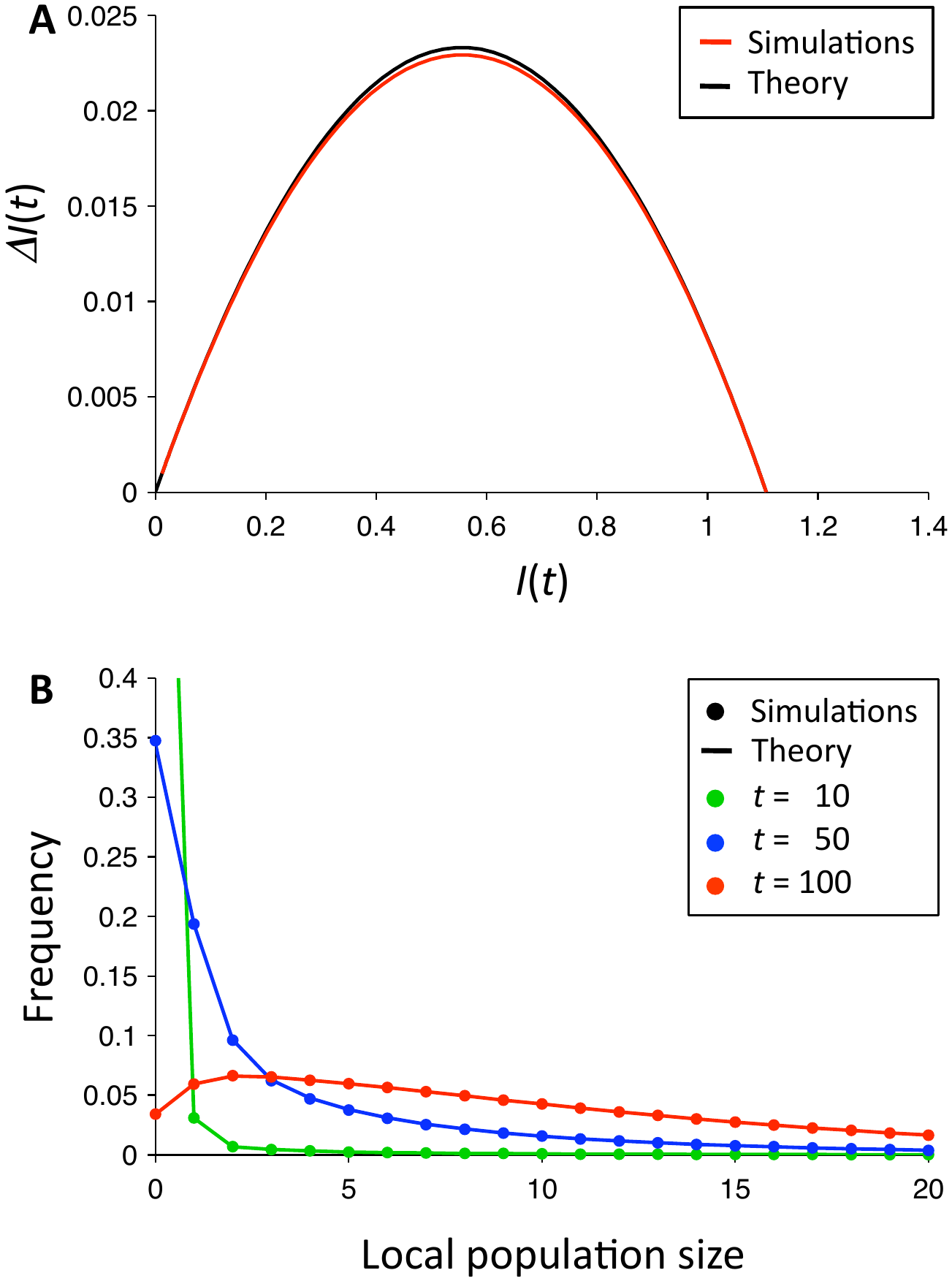}
   \caption{\label{fig:3}
Comparison of theory (\ref{app:B}) to simulations of Eq~\eqref{eq:6}. Panel A compares the change in dispersal rate, $\Delta I$, as a function of $I$, from the simulations (red line) to the theory (black line). Panel B compares the distribution of local population sizes from the simulations (points) to the theoretical predictions (lines), at three different points in time ($t=$10, 50, 100). Parameter values: $R=1.2$, $\alpha=0.01$, $m=0.1$, $k_{\rm D}=1$, $k_{\rm E}=10$ (environmental recruitment stochasticity).}
\end{figure}

Thus we can reconstruct the full distribution of population sizes as a function of $I$, and thereby also estimate $\Delta I$, giving a closed dynamics for $I$. Figure \ref{fig:3}A shows $\Delta I$ as a function of $I$ from simulations (blue line) and from theory (red line). In this example, several higher eigenmodes each give small but significant contributions to $\Delta I$, and are tied (``slaved'') to the first eigenmode. By taking these higher eigenmodes into account, we obtained a very close match between the simulations and the theory. We note that the theory is exact at $I=0$ (trivially) and in the stable steady state (see \ref{app:B} in the ESM). In panel B we show how the distribution of local population sizes evolves as the metapopulation recovers from initially very small local population sizes. The theory (lines) matches the distributions obtained from simulations (dots) very well. 

Thus, the adiabatic solution to the system provides us with the ability to predict the detailed state of the metapopulation based on the effect of dispersal on the local dynamics. From the predicted states, it is the possible to estimate rate of local extinctions, in other words the strength of the rescue effect. This approach has the important advantage of being derived from the full stochastic description of the system (in other words, the rescue effect emerges mechanistically from the interaction of the local stochastic populations). 

\subsection*{Understanding the slow mode in biological terms}


While the separation in fast and slow modes detailed in the previous section is accurate and mathematically tractable, the formulation of slow mode reconstructed above is remains difficult to interpret in biological terms. At the expense of losing some accuracy, we can express the dynamics in terms of the demographic response of a single patch to a constant influx of dispersing individuals. Rather than a full spectral analysis of the dynamics, we express the slow mode as a linear combination of eigenvectors of the matrix $\ma{A}$, and keep only the terms corresponding to the two eigenvalues closest to zero (one eigenvalue is zero, see \ref{app:C} in the ESM for details). This is equivalent to assuming a separation of time scales in the local population dynamics \citep{drechsler1997}. The coefficients of the two eigenvectors are uniquely determined by the constraints that the distribution of patch sizes must sum to unity, and that the rate of dispersal $I$ must match a given value. Putting it all together, we find a closed equation for the change in dispersal rate, $I$, from one generation to the next, $\Delta I$:
\begin{equation}\label{eq:7}
\Delta I = \big( D(I) - I \big)\big( -\lambda_2(I) \big) \, ,
\end{equation}
where $D(I)$ is the average number of individuals dispersing from a population with a stable stream of $I$ individuals (on average, Poisson distributed) arriving to the population each generation, and $\lambda_2(I)$ is the smallest non-zero eigenvalue of $\ma{A}$ in Eq.~\eqref{eq:6}. The first factor in Eq.~\eqref{eq:7}, $D(I)-I$, expresses the imbalance between dispersal from the population and immigration into the population at the local steady state. The second factor in Eq.~\eqref{eq:7}, $-\lambda_2(I)$, reflects the elasticity of $\Delta I$ with respect to this imbalance (i.e. how quickly the demographics of the local populations respond to the imbalance). We note that because single-population dynamics with immigration and dispersal is a Markov process with a unique steady state, the elasticity is always positive. Therefore, the dispersal rate (and hence the mean population size in the metapopulation) increases when the imbalance (the first factor) is positive and decreases when it is negative (reflecting whether patches are sources or sinks for the dynamics, on average). The magnitude of the change in dispersal rate is affected by both factors. Numerically, Eq.~\eqref{eq:7} gives solutions that are qualitatively similar to the slow-mode theory and the simulations (Fig.~\ref{fig:S2} in the ESM). The main difference between the simplified and full reconstruction is the magnitude of $\Delta I$: especially, the sign of $\Delta I$ is still given by the first term in Eq.~\eqref{eq:7}, the imbalance of dispersal and immigration in the local population model (because both theories are exact when $\Delta I=0$).

It is instructive to compare our approximate one-dimensional dynamics to the classical equations in Levins' model. In the simplest version, Levins' equation predicts a simple quadratic relationship between a variable (traditionally the fraction of occupied patches) and its change from one generation to the next:
\begin{equation}\label{eq:8}
\Delta p = c p (p^* - p) \, .
\end{equation}
Here $p^*$ is fraction of occupied patches at the global steady state, and $c$ is the colonisation probability of empty patches. \citet{eriksson2012} showed that under certain circumstances (specifically, when the metapopulation is on the brink of extinction), a Levins-like dynamics may emerge because the distribution of population sizes in occupied patches has a rigid shape, so that only the fraction of occupied patches $p$ changes. In this case, the immigration rate $I$ and the mean population size $\bar{n}$ are both proportional to $p$ (because all three quantities are linear combinations of the underlying distribution of population sizes), and therefore they too have Levins-like dynamics (but with different coefficients in Eq.~\eqref{eq:8}).

In the current model, by contrast, the shape of the distribution of population sizes changes markedly during recovery to the global steady state (Fig.~\ref{fig:1}). As expected, the fraction of occupied patches shows very strong deviations from Levins' equation caused by the rescue effect (Fig.~\ref{fig:S3}A in the ESM, where the expectation from the Levins' equation would be a constant line given by $c$). Surprisingly, however, we find that $I$ and mean population size ($\bar{n}$) both follow rather closely the predictions from Levins equation (Fig.~\ref{fig:S3}B,C) in the ESM). In order to understand this apparent contradiction, we need to consider the dispersal-immigration imbalance ($D(I)-I$) and its elasticity (given by the factor $-\lambda_2(I)$) in Eq.~\eqref{eq:7}. It turns out that the changes in elasticity for different values of $I$ (which are not accounted for by the Levins' equation) are counterbalanced by corresponding changes in the dispersal-immigration imbalance (Fig.~\ref{fig:S3}D in the ESM). Repeating the analysis for the model with environmental stochasticity on survival gave qualitatively similar results (Fig.~\ref{fig:S4} in the ESM).

\section*{The rescue effect in action}

\begin{figure}[tp]
   \includegraphics[width=0.9\textwidth]{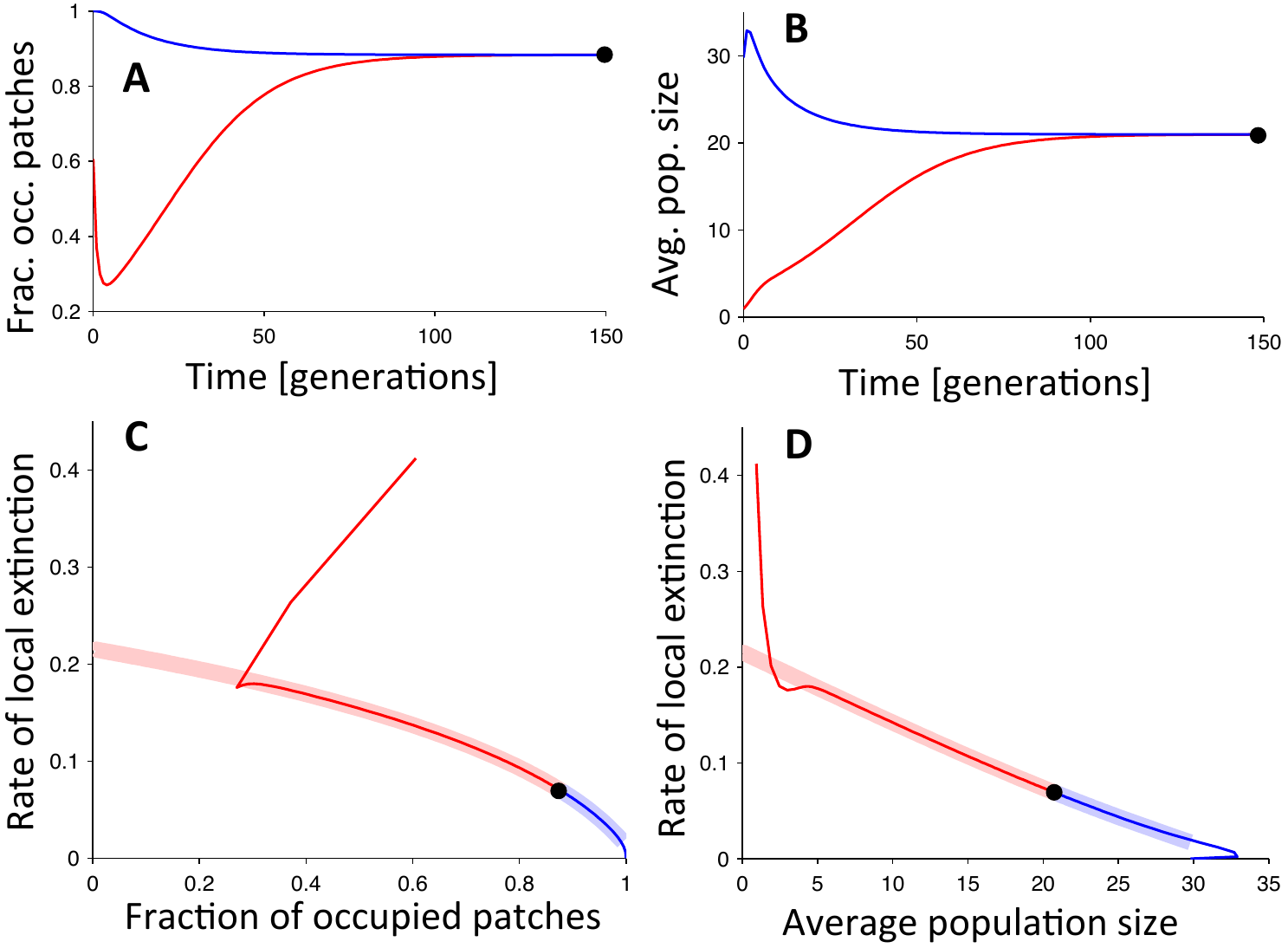}
   \caption{\label{fig:4} 
Time evolution of the metapopulation [Eq.~\eqref{eq:6}] started from above (blue) or below (red) the stable state (the black dot in each panel). Shown are the fraction of occupied patches (A) and the mean population size (B). Probability of local extinction as a function of the fraction of occupied patches (C), and as a function of mean local population size (D) both indicate a strong rescue effect on the persistence of individual populations (thin lines). After a short transient (approximately 5 generations) the trajectories approach the slow mode (thick shaded lines). Parameters are: $R=1.5$, $\alpha=0.01$, $m=0.02$, $k_{\rm D}=0.5$, $k_{\rm E}=5$ (environmental recruitment stochasticity).}
\end{figure}

We explored the behaviour of the metapopulation after a perturbation in the form of a reduction in the population size of all patches (red lines in Fig.~\ref{fig:4}A and B). Because environmental survival stochasticity gave qualitatively similar results, we focus on environmental recruitment stochasticity in Fig.~\ref{fig:4}.  The fraction of occupied patches (Fig.~\ref{fig:4}A) displayed a non-monotonic transient during recovery, with an initial loss of a number of patches due to stochastic extinctions of the diminished populations followed by a progressive recolonisation from the surviving patches. The average population size of occupied patches, on the other hand, recovered monotonically towards the equilibrium value (Fig.~\ref{fig:4}B). The behaviour was different when the system was perturbed by increasing population sizes in all patches. In this case both the fraction of occupied patches and mean population size returned monotonically to their equilibrium values (blue lines in Fig.~\ref{fig:4}A and B).

Using our ability to predict the rate of local extinctions from the distribution of patch sizes (as detailed in \ref{app:B} in the ESM), we explored the probability of extinction for any given patch following perturbations. This allows us to unravel the mechanisms that force the system to return to its steady state. In the absence of a rescue effect, there should be no relationship between local extinction and the fraction of occupied patches. However, we observed a clear inverse relationship between the probability of patch extinction and the fraction of occupied patches (Fig.~\ref{fig:4}C). We compared this to the relationship expected from the slow mode (thick shaded line in Fig.~\ref{fig:4}C), and found that the dynamics very quickly converges to this line, despite starting off the slow mode. We also found a steep inverse relationship between the probability of patch extinction and mean population size (Fig.~\ref{fig:4}D), and also for this relationship a quick convergence to the slow mode. These relationships reflect both the rescue effect and the increasing resilience to demographic stochasticity of larger populations. Thus, in our model, the rescue effect emerges mechanistically from the interaction of the stochastic dynamics of individual populations.

\begin{figure}[tp]
   \includegraphics[width=0.6\textwidth]{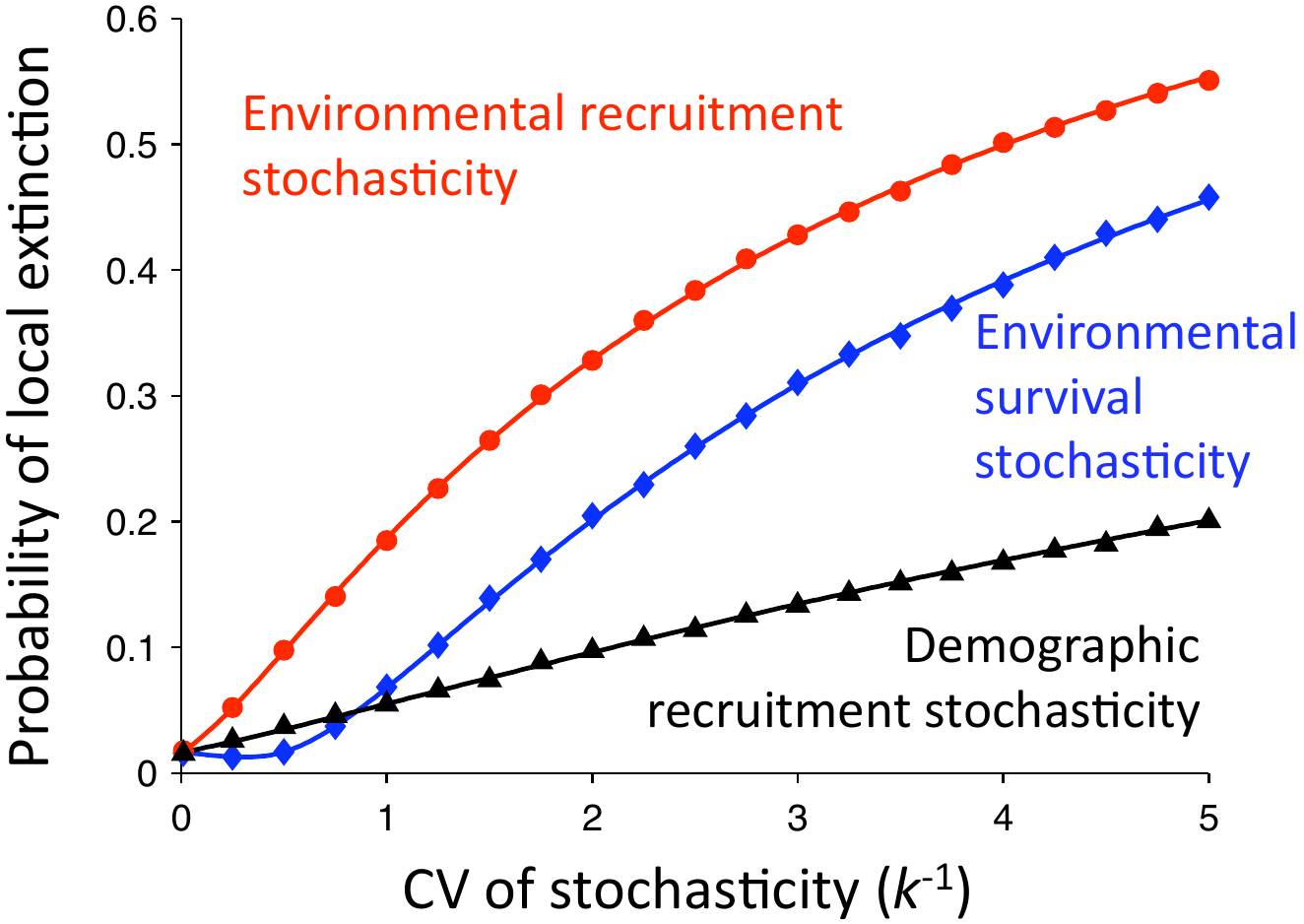}
   \caption{\label{fig:5}
The effect of different types of stochasticity on the probability of local extinction (symbols are simulations ($N=900$ patches) and lines are the theory (\ref{app:B})). We show curves for demographic recruitment stochasticity (black); environmental recruitment stochasticity (red); and environmental survival stochasticity (blue). Each type of stochasticity is considered separately, and we condition on having mean local population size equal to 10. The level of stochasticity (demographic or environmental) is quantified as the coefficient of variation (CV) of the individual reproductive number ($R_i$). Parameters are $R=1.2$, $\alpha=0.01$, and $m=0.1$.}
\end{figure}

Compared to models in which the rescue effect is represented by an arbitrary function linking extinction rates with the proportion of occupied patches, the mechanistic nature of our framework allows us to investigate the relative roles of demographic and environmental stochasticity in affecting the strength of the rescue effect. We considered the effect of each type of stochasticity separately by setting the strength of the other kinds to zero (Fig.~\ref{fig:5}). As we were interested in the rescue effect, we conditioned our simulations to have the same mean within-patch occupancy (ten individuals, thus standardising the metapopulation state). While increasing either type of stochasticity had a negative effect on patch persistence, environmental stochasticity had a much stronger impact on the rescue effect than the demographic stochasticity (Fig.~\ref{fig:5}). This difference arises because environmental stochasticity tends to affect all individuals within the same patch simultaneously, thus giving little room for the rescue effect to come into play. On the other hand, demographic stochasticity will only affect part of the population, thus leaving small populations that can be rescued. Among the two types of environmental stochasticity, recruitment stochasticity had a larger negative effect than survival stochasticity. Interestingly, very low levels of environmental survival stochasticity led to a slight decrease in extinction probability (a phenomenon also observed by \citet{chesson1982}).
 
\section*{Discussion} 
\label{sec:discussion}

Our analytical approach provides a powerful way to explore the link between local and metapopulation dynamics. While the distribution of local population sizes in a metapopulation undergoes complex changes while the population approaches the global steady state, it was possible to find a close approximation to the full population dynamics using the instantaneous dispersal rate as the dynamic variable. Simulations showed that our analytical approach is perfectly adequate in describing metapopulations with localised dispersal, as long as the range of dispersal extends beyond the immediate neighbours. This robustness to the exact scale of dispersal makes our approach widely applicable, providing a convenient alternative to individual-based simulations when investigating metapopulations with explicit local dynamics. An important property highlighted by our framework is the emergence of the rescue effect from the interaction of local dynamics mediated via dispersal.

Our framework makes it possible to quantify the rescue effect, as determined by the local dynamics, in models where the strength of the effect has to be defined a-priori. For example, when investigating extinction debt using the framework developed by \citet{ovaskainen2002}, the rescue effect is represented by introducing non-linearities in the colonisation and extinction functions. The shape and scale of such non-linear functions could only be quantified if several long time-series collected under similar circumstances were available, a luxury that is usually not available. Within our framework, the same feat can be achieved simply by taking into consideration the local population dynamics, something that is often studied and quantified by ecologists. A further advantage is that our approach also allows us to investigate how changing the local dynamics might affect the whole metapopulation.

Local stochasticity, and in particular its environmental component, was shown to play an important role in determining the metapopulation dynamics, especially during the recovery from perturbation. Local environmental stochasticity not only had a direct effect on the local population size, but also had the potential to negate the rescue effect from the rest of the metapopulation. These results are in line with much of the literature on this topic \citep{lande1998,melbourne2008,higgins2009,saether1999,casagrandi1999}, and stem from the fact that local environmental stochasticity affects all individuals within the same population simultaneously, and thus has a much stronger effect on the persistence time of the population compared to demographic stochasticity. Global environmentally driven fluctuations (affecting all populations simultaneously), which were not modelled in this paper, have the potential of also affecting recolonisation rates due to synchrony within the metapopulation \citep{higgins2009,casagrandi1999}. It will be interesting in the future to explore the link between local demography, global environmental stochasticity, and metapopulation persistence time.

The key take-home message of our analysis is that local population dynamics can play a very important role in determining the metapopulation behaviour, directly as well as via the rescue effect. We have provided a convenient analytical framework that allows this role to be investigated, and found our approach to be surprisingly robust to the spatial details to dispersal (except for very localised movement to adjacent patches). Interestingly, the dynamics of the metapopulation can be reconciled with a Levins-like dynamics with appropriately modified rates. This provides an opportunity to establish which local dynamics are compatible with previous models based on Levins' equation and its generalisations.

\subsection*{Acknowledgements}

BM gratefully acknowledges financial support by Vetenskapsr\aa{}det, by the G\"oran Gustafsson Foundation for Research in Natural Sciences and Medicine, and by the Centre for Evolutionary Marine Biology at Gothenburg University.



\bibliography{rescue_effect} 

\newpage

\begin{figure}[tp]
   \includegraphics[width=0.6\textwidth]{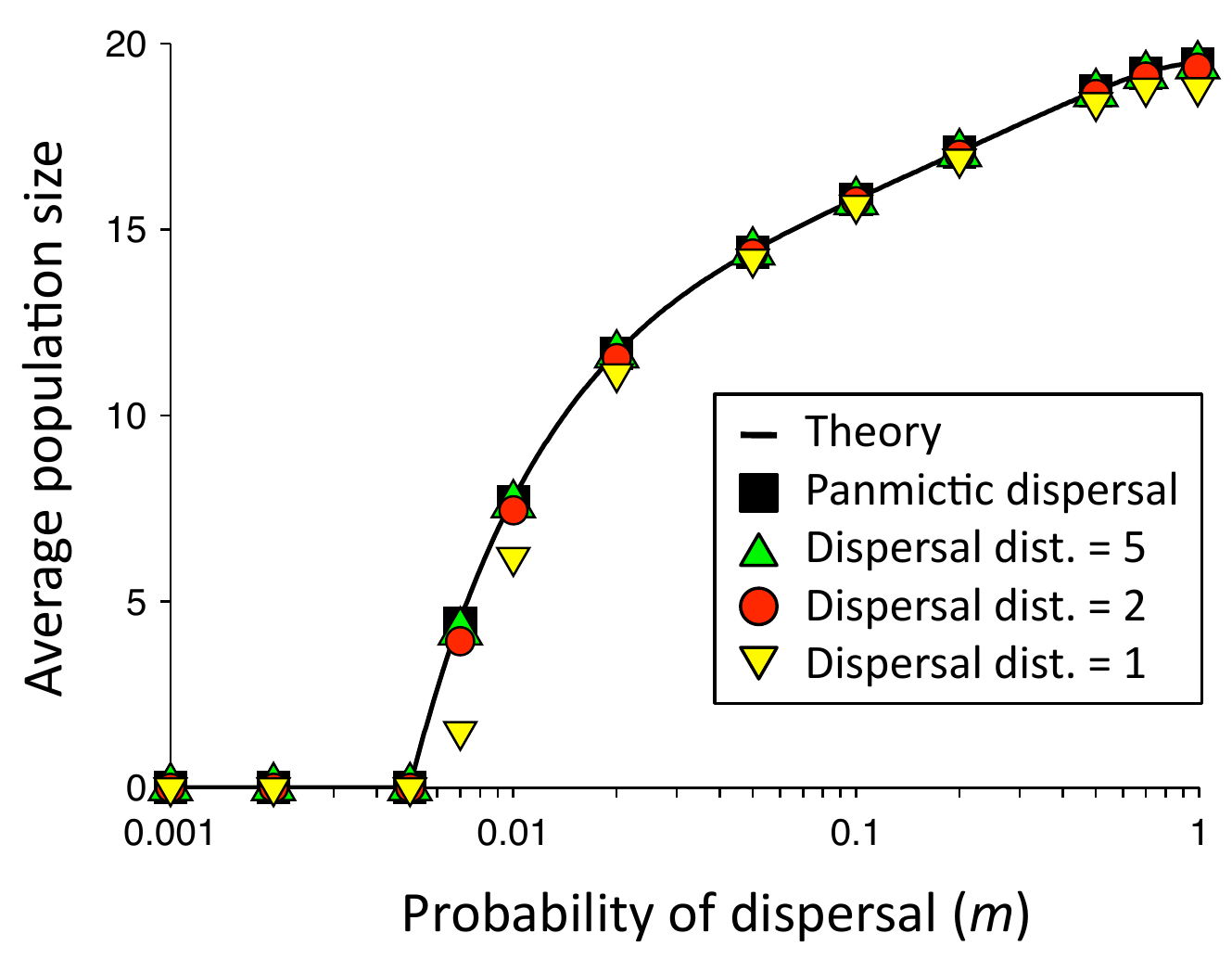}
   \caption{\label{fig:2}
   The effect of the dispersal distance and dispersal rate on the mean stationary population size in a spatially explicit model (20 by 20 patches in a square lattice). Simulations of panmictic dispersal (squares, all patches equally likely to receive a dispersing individual) are in excellent agreement with the theory for infinite number of patches (Eq. \ref{eq:6}, line). Simulations of dispersal with random direction and exponentially distributed distance (circles and triangles) show that as soon as the mean dispersal distance exceeds the distance between demes, the theory is in very good agreement with the simulations. Shown for $R=1.5$, $\alpha=0.02$, $k_{\rm D}= 1$, and $k_{\rm E} =10$ (environmental recruitment stochasticity). Each simulation data point was estimated from 10,000 independent measurements.}
\end{figure}

\clearpage

\section*{\large Electronic Supplementary Material}
\setcounter{page}{1}

\section*{Contents}

\begin{tabular}{lcr}
Appendix A & \hspace{13cm} & 1\\
Appendix B &  & 2\\
Appendix B &  & 3\\
Figure S1 &  & 5\\
Figure S2 &  & 6\\
Figure S3 &  & 7\\
Figure S4 &  & 8\\ 
\end{tabular}

\section*{Supplementary Text}

\renewcommand\thesubsection{Appendix \Alph{subsection}}
\numberwithin{equation}{subsection}
\subsection{Individual-based simulations}\label{app:A}

In our simulations of the model described in the main text, the populations evolve as follows. First, for each population, we used inverse-transform sampling to generate the number of individuals in each population following recruitment and density-dependent mortality from the distribution of possible population sizes given by Eq.~\eqref{eq:1} in the main text.

Second, each individual disperses with probability $m$, otherwise remaining in its native population. In the case of the unstructured metapopulation, this is implemented by first moving all dispersing individuals to a common pool, and then moving each dispersing individual to a randomly chosen target population.

The simulations for the spatially explicit metapopulation (Fig. \ref{fig:2} in the main text) differ in the way the destination of the dispersers is chosen. In this case the populations are located in a square grid, where the sides of the squares are of size unity. Consider a dispersing individual leaving a population located in a particular square. Its destination is selected by the following procedure. First, a dispersing distance is chosen by generating an exponentially distributed random number such that the mean distance parameter is $d_{\rm dispersal}$. Second, a direction is chosen by a randomly generated angle. Third, from the centre of the square a line is drawn with the previously chosen length and angle, and the dispersing individual is placed in the population located at the end of the line (using periodic boundaries where necessary).

\subsection{Finding the slow mode}\label{app:B}
\renewcommand{\theequation}{B.\arabic{equation}}

The dynamics of the model (Eq.~\ref{eq:6} in the main text) is nonlinear in $\ve f$, where $\ve f$ is the distribution of local population sizes in vector form, but the nonlinear term depends only on the dispersal rate $I$. In this appendix we show how to find the local slow mode for a given value of $I$ by analysing fast and slow modes of the dynamics. Using Eqs.~(\ref{eq:1},\ref{eq:2},\ref{eq:4}) in the main text we can write the
dispersal rate as $I=\ve v\tr \ve f$ where the vector $\ve v$ has elements $v_j = \sum_{i=1}^{\infty}m i P_{ij}$ and ``\textsf{T}'' denotes transpose. The fast and slow modes are found by analysing the eigenmodes of the dynamics linearised around the current population size distribution $\ve f$. This is determined by the Jacobian matrix $\ma {J}$:
\begin{equation}\label{eq:B1}
\ma J(\ve f) = \frac{\partial \Delta \ve f}{\partial \ve f} = \frac{\partial}{\partial \ve f}\Big[ \ma A\big(\ve v\tr \ve f\big)\ve f \Big] = \ma A(I) + \ma A'(I) \ve f \ve v\tr \, ,
\end{equation}
where $\ma A'$ is the derivative of the matrix $\ma A$ with respect to $I$.

Our task is to reconstruct $\ve f$ conditional on the immigration rate $I$. To this end, we express the matrix $\ma A$ and the population size distribution $\ve f$ in Eq.~\eqref{eq:B1} in terms of the left and right eigenvectors of $\ma J$ (denoted $\ve L_i$ and $\ve R_i$, respectively, for the $i$th eigenvalue). For simplicity, we number the eigenvalues $\lambda_i$ in order of increasing magnitude ($|\lambda_1|<|\lambda_2|<|\lambda_3|<\dotsb$) and take left and right eigenvectors to be bi-orthonormal. Expanding $\ve f$ in the right eigenvectors of $\ma A$, we obtain
\begin{equation}\label{eq:B2}
\ve f = \sum_{i=1}^{\infty}c_i \ve R_i \, .
\end{equation}
We note that the constant vector of ones is the left eigenvector for $\ma J$ corresponding to eigenvalue zero (this simply reflects that $\ve f$ is normalised to unity). Hence, the corresponding eigenvector $\ve R_1$ enters in $\ve f$ with coefficient $c_1 = 1$. The remaining coefficients are determined from the constraint $I=\ve v\tr \ve f$,
\begin{equation}\label{eq:B3}
I = \ve v\tr \ve f = \sum_{i=1}^{\infty}c_i\ve v\tr \ve R_i \, ,
\end{equation}
and the adiabatic principle that the change in the population size distribution along the third and higher eigenmodes should vanish, i.e.
\begin{equation}\label{eq:B4}
0 = \ve L_i\tr \Delta \ve f = \sum_{j=i}^{\infty} \ve L_i\tr \ma A \ve R_j c_j\,, \quad \text{for $i \geq 3$.}
\end{equation}
Together, these constraints give a linear system of equations from which the coefficients $c_2$, $c_3$, $c_4$, $\dotsc$ that we solve numerically (in practice we normally truncate the system, taking only the $c_1$, $\dotsc$, $c_{10}$ into account). This ties (`slaves') the coefficients of the higher-order modes to the first and second eigenmodes.

Since the Jacobian $\ma J$ depends explicitly on the population-size distribution $\ve f$, and not only on the rate of immigration $I$, we must find $\ve f$ in a self-consistent manner. In the first iteration we calculate $\ma J$ from Eq.~\eqref{eq:B1} using the given value of $I$ and take $\ve f$ to be zero, and use Eqs.~\eqref{eq:B3} and \eqref{eq:B4} to determine the coefficients $c_i$ and thereby $\ve f$. We then repeat these steps, using the latest estimate of the population-size distribution as input to each iteration, until convergence. Although there is in general no guarantee of convergence, we have found this approach to be quite stable for a wide range of parameters (5-10 iterations have been sufficient for the parameter combinations explored in this paper).

\subsection{Simple approximation}\label{app:C}
\renewcommand{\theequation}{C.\arabic{equation}}

In this appendix we derive a simple, explicit approximation for the dynamics of the dispersal rate $I$. As in \ref{app:B}, we use spectral analysis to identify fast and slow mode. Here, however, we will make two simplifying assumptions: first, that the second term in Eq.~\eqref{eq:B1} is small enough that $\ma A$ can be used to approximate $\ma J$; second, that it is sufficient to keep only the first and second eigenvalues ($c_1$ and $c_2$) in the expansion of $\ve f$ [Eq.~\eqref{eq:B2}], setting all other coefficients to zero. Both of these assumptions are valid if, for example, there is a separation of time scale between the local population dynamics and the global dynamics of $I$, but also when this is not the case the approximation can be used qualitatively.

Using these assumptions, we can write the dynamics of $I$ as
\begin{equation}\label{eq:C1}
\Delta I = \ve v\tr \Delta \ve f = \ve v\tr \ma A(I) \big( c_1\ve R_1 + c_2 \ve R_2 \big) =
 c_2\lambda_2 \ve v\tr \ve R_2  \, ,
\end{equation}
where we have used that $\ve R_1$ and $\ve R_2$ are eigenvectors of $\ma A$ and that $c_1=1$ (since all columns of $\ma A$ sum to zero, we find again that $c_1=1$ for all $I$). Using equation \eqref{eq:B3}, we obtain
\begin{equation}\label{eq:C2}
I = \ve v\tr \ve f = D(I) + c_2 \ve v\tr \ve R_2 \, ,
\end{equation}
where $D(I) = \ve v\tr \ve R_1$ can be interpreted as the rate of dispersal from a population in the
steady state determined by the immigration rate $I$. Thus, solving Eq.~\eqref{eq:C2} for $c_2$ and inserting it
into Eq.~\eqref{eq:C1} we obtain Eq.~\eqref{eq:7} in the main text. In addition, the population size distribution $\ve f$
predicted from the immigration rate $I$ is given by
\begin{equation}\label{eq:C3}
\ve f = \ve R_1 + \frac{I - D(I)}{\ve v\tr \ve R_2}\ve R_2 \, .
\end{equation}
We conclude this section with a couple of comments on the relation between the simple approximation and the full adiabatic treatment in \ref{app:B}. First, because of the constraint $I=\ve v\tr \ve f$ and Eq.~\eqref{eq:B2}, the coefficients $c_2$, $c_3$, $\dotsc$ are all approximately proportional to $I - \ve v\tr \ve R_1$. Because the local steady-state changes only slowly with $I$, this factor is very close to $I- D(I)$, the corresponding factor in the simple approximation. Thus, also the full adiabatic treatment supports the interpretation, from the simple approximation, that the metapopulation dynamics is equivalent to a single population that is continuously approaching the steady state given by the current immigration rate, but prevented from reaching it because the immigration rate changes in the very process, until the metapopulation reaches the global steady state.



\newpage

\section*{Supplementary Figures}

\renewcommand{\thefigure}{S\arabic{figure}}
\setcounter{figure}{0}

\begin{figure}[h]
   \includegraphics[width=0.6\textwidth]{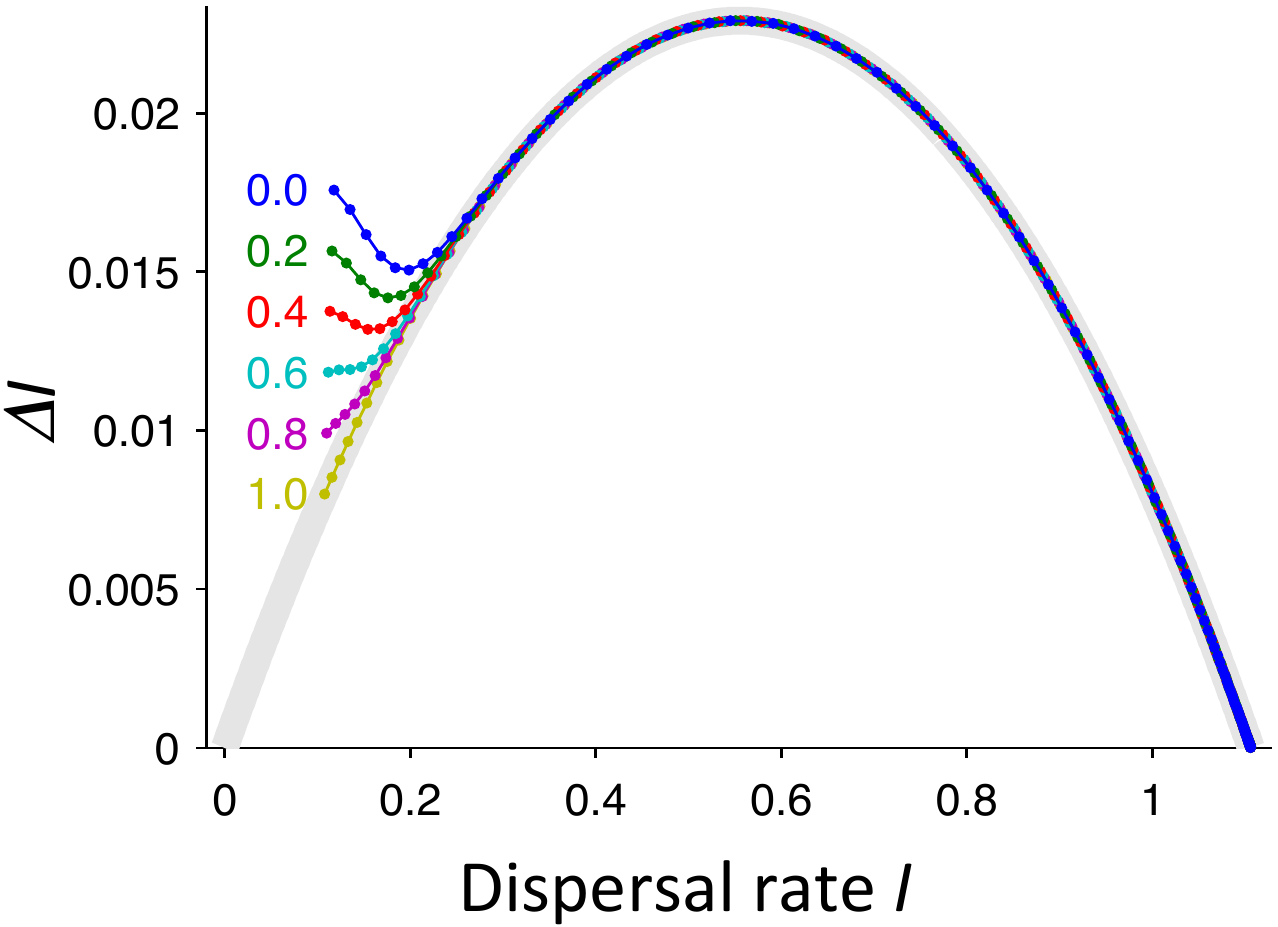}
   \caption{\label{fig:S1}  
Illustration of convergence to the slow mode. We started simulations (Eq.~\eqref{eq:6}, dotted lines, one dot for each time step) from distributions of patch occupancy that are linear mixes of two distributions with the same mean local population size (equal to one), such that 1 is the slow-mode distribution, and 0 is the Poisson distribution. Because of linearity, all intermediate distributions have the same mean population size. The grey thick line is the relationship between dispersal rate $I$ and $\Delta I$ predicted by theory. Even when the simulations start far away from the slow mode, the trajectories converge to the slow mode within a few time steps. Parameter values: $R=1.2$, $\alpha=0.01$, $m=0.1$, $k_{\rm D}=1$, $k_{\rm E}=10$ (environmental recruitment stochasticity).}
\end{figure}

\begin{figure}[tp]
   \includegraphics[width=0.6\textwidth]{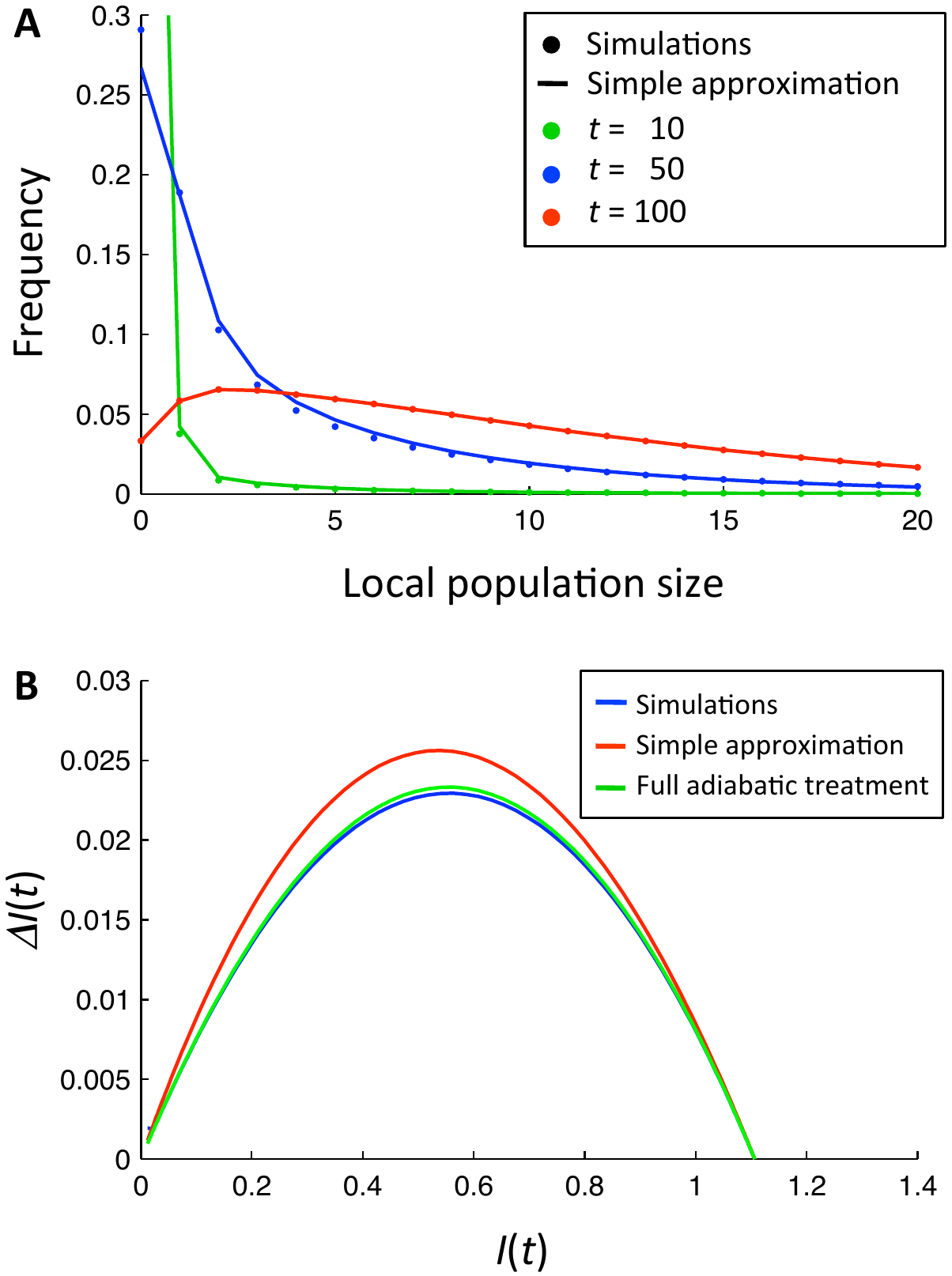}
   \caption{\label{fig:S2}  
Comparison of the simple approximation (Eq.~\eqref{eq:7}) to simulations (Eq.~\eqref{eq:6}) and the full adiabatic treatment (\ref{app:B}). Panel A compares the distribution of local population sizes from the simulations (points) to the simple approximation, at three different points in time ($t=$10, 50, 100). Panel B compares the change in dispersal rate, $\Delta I$, as a function of $I$, in the simulations to the simple approximation and the full theory. Parameter values: $R=1.2$, $\alpha=0.01$, $m=0.1$, $k_{\rm D}=1$, $k_{\rm E}=10$ (environmental recruitment stochasticity).}
\end{figure}

\begin{figure}[tp]
   \includegraphics[width=0.75\textwidth]{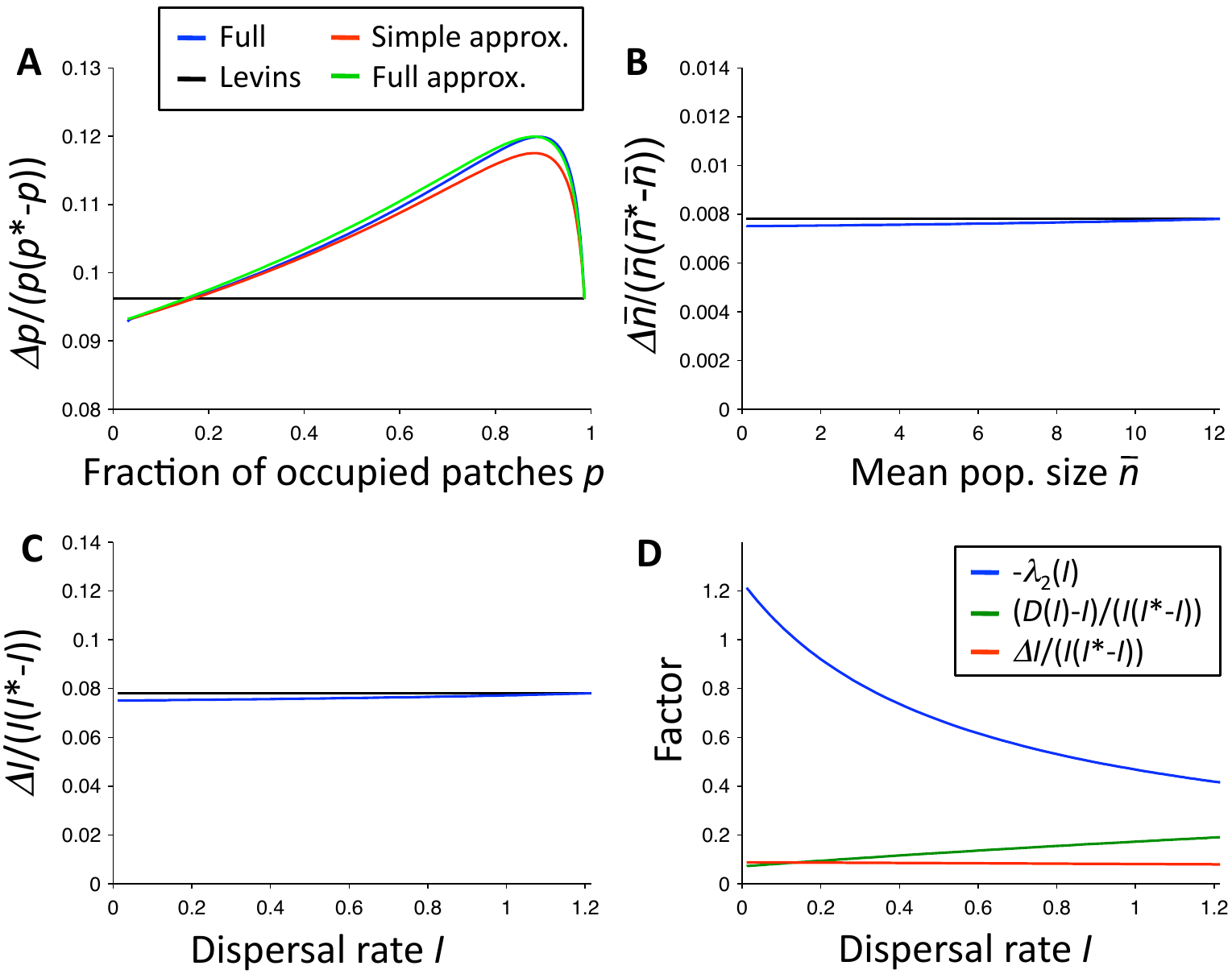}
   \caption{\label{fig:S3}  
Comparison of our model to Levins' model. A: In Levins' model the change in fraction of occupied patches ($\Delta p$) is proportional to $p(p^*-p)$ where $p^*$ is the value of $p$ in the stationary state of the metapopulation (c.f. Eq. \eqref{eq:8} in the main text). According to Levins' simple equation, in the absence of a rescue effect, $\Delta p/p(p^*-p)$ is constant (black line). Deviations of $\Delta p/p(p^*-p)$ from the constant value reveal how the rescue effect emerges from the local population dynamics in our model. We show the full deterministic dynamics (Eq.~\eqref{eq:6}, blue line) for a trajectory starting from mean population size 0.1, and the corresponding curves for the slow-mode theory (\ref{app:B}, green line) and the simple approximation (Eq.~\eqref{eq:7}, red line). We observe a strong rescue effect and excellent agreement between the analytical approximations and the full dynamics. In contrast, the corresponding analysis for the mean population size ($\bar{n}$, panel B) and dispersal rate ($I$, panel C) shows that the dynamics of these observables is very close to Levins' model (after a short transient). D: The two factors determining the change in dispersal rate, $\Delta I$, in the simple approximation, as a function of $I$. As can be seen in panel C and from the red line, $\Delta I$ is approximately proportional to $I(I^*-I)$, where $I^*$ is the steady-state dispersal rate. This would seem to imply an absence of rescue effect, but plotting the two factors in the simple approximation of $\Delta I$ [Eq.~\eqref{eq:6}] separately reveals that the effect of an increasing trend in the excess emigration rate ($D(I)-I$, green line) is offset in $\Delta I$ by a corresponding decreasing trend in the elasticity ($-\lambda_2(I)$, blue line). Parameter values: $R=1.2$, $\alpha=0.01$, $m=0.1$, $k_{\rm D}=1$, $k_{\rm E}=10$ (environmental recruitment stochasticity).}
\end{figure}

\begin{figure}[tp]
   \includegraphics[width=0.75\textwidth]{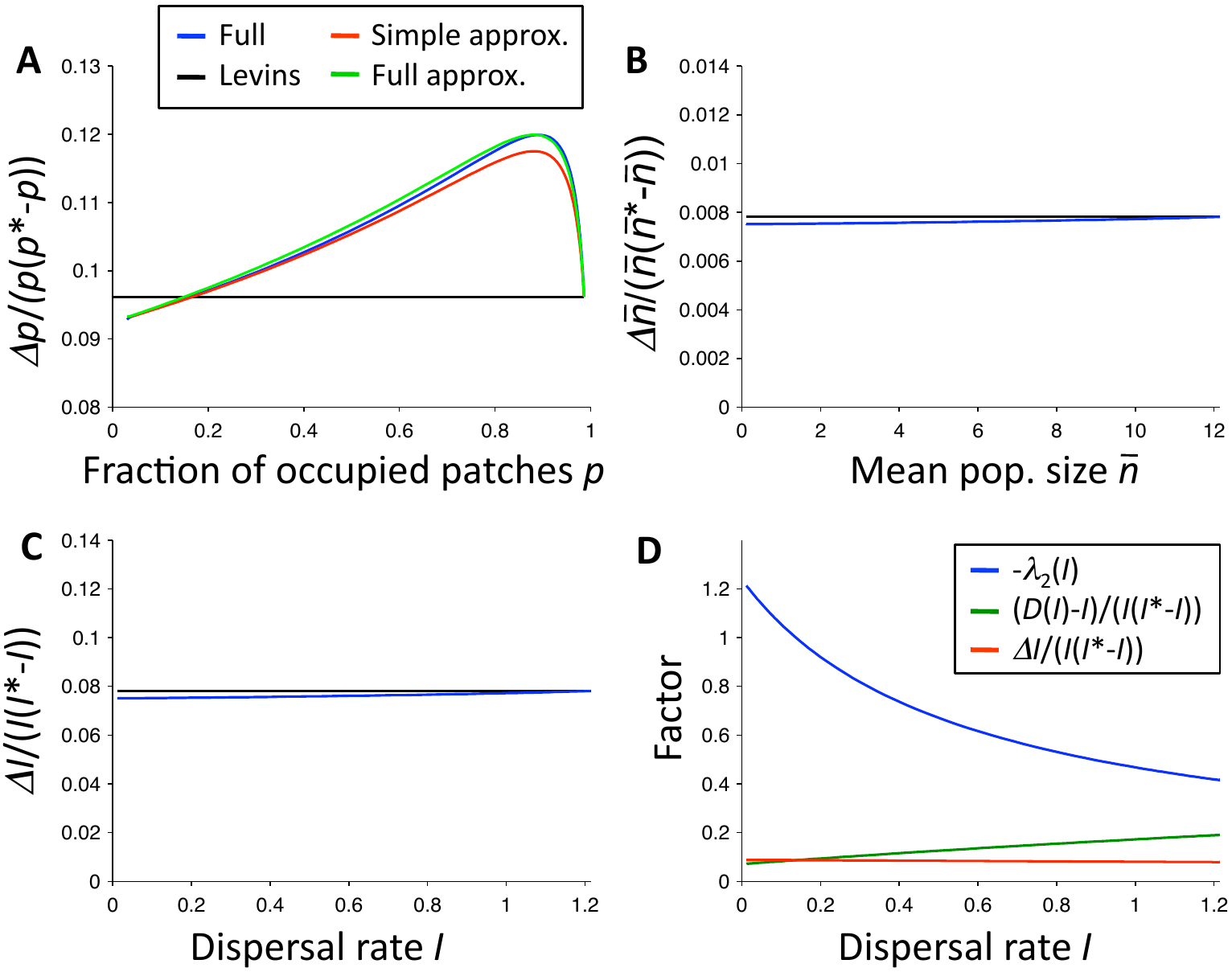}
   \caption{\label{fig:S4}  
Comparison of our model to Levins' model for environmental survival stochasticity [Eq. \eqref{eq:1}b], analogous to Figure \ref{fig:S3}. A: Comparison to Levins' model for the fraction of occupied patches. B,C: Levins' model is a good fit of the dynamics of the mean population size and of the dispersal rate. D: The two factors determining the change in dispersal rate, $\Delta I$, in the simple approximation, as a function of $I$. Parameter values: $R=1.2$, $\alpha=0.01$, $m=0.1$, $k_{\rm D}=1$, $k_{\rm E}=10$.  For a more full explanation of each panel, see caption of Figure \ref{fig:S3}.}
\end{figure}

\end{document}